\documentclass[aps,prl,showpacs,twocolumn,superscriptaddress,letterpaper,longbibliography]{revtex4-1}
\usepackage{graphicx}
\usepackage{dcolumn}
\usepackage{bm}
\usepackage{amsmath, amssymb}
\usepackage{epstopdf}
\usepackage{color}
\usepackage[normalem]{ulem}
\usepackage{multirow}
\usepackage{enumitem}
\usepackage{comment}
\usepackage{hyperref}
\usepackage{cleveref}
\usepackage{lineno}
\usepackage{setspace}
\usepackage{blindtext}

\hypersetup{breaklinks = true, colorlinks = true, citecolor = blue, linkcolor = blue, urlcolor = blue}

\begin{document}

\title{A=3 \texorpdfstring{$(e,e')$ $x_B \ge 1$}{(e,e') xB >= 1} cross-section ratios and the isospin structure of short-range correlations}

\date{\today}

\newcommand{\MIT}{Massachusetts Institute of Technology (MIT), Cambridge, MA, 02139, USA}
\newcommand{\GWU}{The George Washington University, Washington, DC, 20052, USA}
\newcommand{\ODU}{Old Dominion University, Norfolk, VA 23529, USA}
\newcommand{\TAU}{Tel Aviv University, Tel Aviv, 6927845, Israel}
\newcommand{\JLab}{Thomas Jefferson National Accelerator Facility, Newport News, VA 23606, USA}
\newcommand{\UW}{Department of Physics, University of Washington, Seattle, WA 98195, USA}
\newcommand{\HUJI}{The Racah Institute of Physics, The Hebrew University, Jerusalem, 9190401, Israel}
\newcommand{\LANL}{Theoretical Division, Los Alamos National Laboratory, Los Alamos, New Mexico 87545, USA}
\newcommand{\PSU}{Department of Physics, The Pennsylvania State University, Norfolk, Virginia, 23529, USA}

\author{A.~Schmidt} \affiliation{\GWU}
\author{A.~W.~Denniston} \affiliation{MIT}
\author{E.~M.~Seroka} \affiliation{\GWU}
\author{N.~Barnea} \affiliation{\HUJI}
\author{D.W.~Higinbotham} \affiliation{\JLab}
\author{I.~Korover} \affiliation{\MIT}
\author{G.A.~Miller} \affiliation{\UW}
\author{E.~Piasetzky} \affiliation{\TAU}
\author{M.~Strikman} \affiliation{\PSU}
\author{L.B.~Weinstein} \affiliation{\ODU}
\author{R.~Weiss} \affiliation{\LANL}
\author{O.~Hen} \affiliation{\MIT}

\begin{abstract}
We study the relation between measured  high-$x_B$, high-$Q^2$, Helium-3 to Tritium, $(e,e')$ inclusive-scattering cross-section ratios and the relative abundance of high-momentum neutron-proton ($np$) and proton-proton ($pp$) short-range correlated (SRC) nucleon pairs in three-body ($A=3$) nuclei.  
Analysis of this  data using a simple pair-counting cross-section model suggested a much smaller $np/pp$ ratio than previously measured  in  heavier nuclei, questioning our understanding of  $A=3$ nuclei and, by extension, all other nuclei.  
Here we examine this finding using spectral-function-based cross-section calculations, with both an \textit{ab initio} $A=3$ spectral function and effective Generalized Contact Formalism (GCF) spectral functions using different nucleon-nucleon interaction models. The \textit{ab initio} calculation agrees with the data, showing good understanding of the structure of $A=3$ nuclei. An 8\% uncertainty on the simple pair-counting model, as implied by the difference between it and the \textit{ab initio} calculation, gives a factor of 5 uncertainty in the extracted $np/pp$ ratio.
Thus we see no evidence for the claimed ``unexpected structure in the high-momentum wavefunction for hydrogen-3 and helium-3''.
\end{abstract}

\maketitle

Physics advances fastest by comparing high-precision experiments with detailed calculations.  
Hard scattering reactions on few-body nuclear systems are accessible to both precision measurements and precision calculations, allowing us to advance our knowledge of nuclear structure and dynamics. 

Recently  Helium-3 and Tritium inclusive $(e,e')$~\cite{Li:2022fhh} and semi-inclusive $(e,e'p)$~\cite{Cruz-Torres:2020uke} scattering measurements were performed to provide new insight into the properties of high-momentum short-range correlated (SRC) nucleon pairs in  $A=3$ nuclei. SRC pairs are fluctuations of strongly-interacting nucleon pairs in nuclei. Their measurable properties, such as isospin structure (i.e., neutron-proton to proton-proton pair ratio, $np/pp$), provide unique insight into the nature of the short-distance nuclear interaction and the behavior of nucleons  at very high momenta~\cite{ciofi15,Hen:2016kwk}.

The measured $(e,e'p)$ cross-sections were compared to modern calculations using \textit{ab initio}  spectral functions (i.e., spectral functions calculated precisely using a specific $NN$ potential)~\cite{Golak:2005iy} for $A=3$, agreeing well up to 450 MeV/c  initial nucleon momentum. This is a remarkable success of modern nuclear theory in extreme conditions of SRC pair formation~\cite{Cruz-Torres:2020uke}. However, the measured $(e,e')$ $^3$H/$^3$He cross-section ratios have so far only been analyzed using a simple SRC pair-counting model~\cite{Li:2022fhh}. This analysis reported a dramatically smaller $np/pp$ SRC ratio than previous $4\leq A \leq 208$ measurements~\cite{piasetzky06,subedi08,korover14,hen14,Duer:2018sxh}, and $3\leq A \leq 40$ calculations~\cite{Cruz-Torres:2019fum}. 
This decrease was attributed to an ``unexpected structure in the high-momentum wave-function for hydrogen-3 and helium-3''~\cite{Li:2022fhh}. As the $A=3$ few-body system has been thoroughly studied both experimentally and theoretically, such a new structure would be most surprising and could  dramatically impact  our understanding of all nuclei. 

This  necessitates a study of the consistency of the measured $(e,e')$ data with modern \textit{ab initio} nuclear structure and reaction theory to determine the accuracy with which it can constrain the SRC isospin structure in $A=3$ nuclei.

Here we compare the measured $(e,e')$ cross-section ratios to different theoretical models with different levels of approximations.
These include calculations using an \textit{ab initio} spectral function~\cite{AttiKaptari:2005} calculated with the AV18 interaction~\cite{wiringa95} and calculations using  effective SRC-pair spectral functions obtained using the Generalized Contact Formalism (GCF)~\cite{Weiss:2015mba,Weiss:2016obx,Weiss:2018tbu,Cruz-Torres:2019fum} with different nucleon-nucleon ($NN$) interaction models~\cite{wiringa95,Wiringa:2002,Piarulli:2016vel,Piarulli:2017dwd,Baroni:2018fdn,Gezerlis:2014,Lynn:2016,Lonardoni:2018prc}.

We find that the full spectral-function calculation describes the measured $(e,e')$ $^3$H/$^3$He cross-section ratios well.  This implies that there is no new structure in the $A=3$ momentum distribution.  This calculation disagrees with the simple pair-counting model by 8\%.   We attribute this  difference primarily to non-SRC-pair contributions to the measured cross-section.
An 8\% uncertainty in the simple pair-counting model causes a factor of five uncertainty in the extracted $np$ to $pp$ SRC pair ratio, eliminating the discrepancy with previous results.

In inclusive electron scattering, an electron with  energy $E_{e}$ scatters off a target, deflecting by an angle $\theta_{e'}$ and emerging with energy $E_{e'}$.
The kinematics is  described in terms of the momentum transfer squared $Q^2 = 4  E_{e}  E_{e'}  \sin^2(\theta_{e'}/2)$ and the Bjorken scaling variable $x_B = Q^2 / (2 m_N \omega)$, where $\omega = E_{e} - E_{e'}$ is the  energy transfer and $m_N$ is the nucleon mass.

At $Q^2 \ge 1.5$ GeV/c$^2$, measured inclusive $(e,e')$  cross-section ratios of nuclei relative to deuterium ``scale" for $1.5\leq x_B \leq 1.8$ ({\it i.e.}, the   ratio is independent of $x_B$)~\cite{frankfurt93,egiyan02,Schmookler:2019nvf}. At this  high $x_B$, the electrons  interact predominantly with high-momentum nucleons~\cite{egiyan02}. Thus the observed scaling is  interpreted as suggesting that  inclusive scattering  proceeds predominantly via the hard breakup of SRC pairs. 

Using this simple reaction picture, early works introduced the SRC cross-section approximation~\cite{Frankfurt81}:
\begin{equation}
    \label{eq:a2}
    \sigma_{eA}(x_B,Q^2) = a_2(A) \cdot \frac{A}{2} \cdot \sigma_{ed}(x_B,Q^2),
\end{equation}
where $a_2(A)$ is  the ratio of the prevalence of SRC pairs in nucleus $A$ relative to the high-momentum fraction of the deuteron momentum distribution, and $\sigma_{ed}$ is the electron-deuteron scattering cross-section. In a high-resolution reaction picture, the latter includes interactions with either the proton or the neutron in the deuteron, i.e. $\sigma_{ed} \approx \sigma_{ep} + \sigma_{en}$, where $\sigma_{ep}$ and $\sigma_{en}$ are the off-shell electron-nucleon cross-sections~\cite{DeForest:1983ahx}.
In this approximation,  $a_2(A) = \sigma_{eA}/\sigma_{ed}$  provides information on SRC pairing probabilities in nuclei.

\begin{figure}[t]
\centering 
\includegraphics[width=0.85\linewidth]{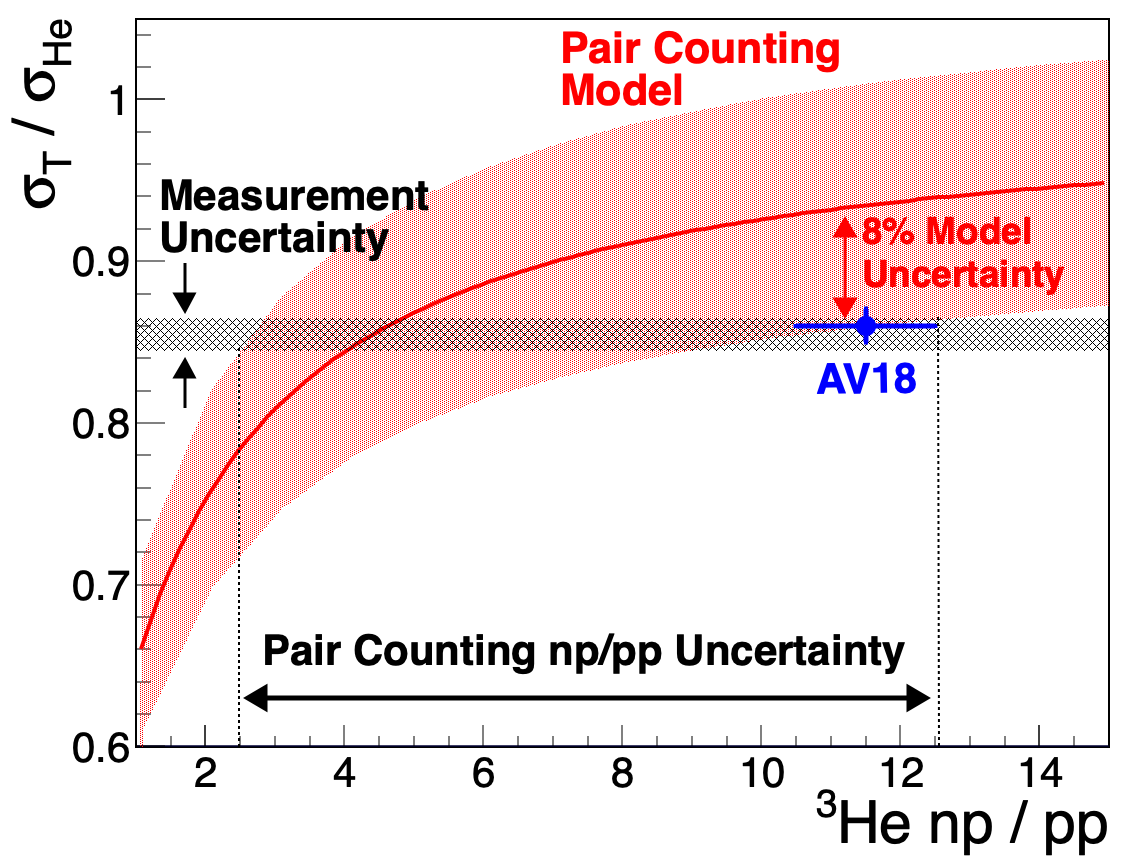}
\caption{
The relation between the average  $^3$H/$^3$He cross-section ratio for $1.4 \le x_B \le 1.7$ and the relative number of $np$ and $pp$ SRC pairs in Helium-3.
The experimental results of Ref.~\cite{Li:2022fhh} are shown by the black band, whose width  equals  the data uncertainty.
The AV18 prediction is shown by the blue data data point using the $np/pp$ SRC-pair ratio from Ref.~\cite{Cruz-Torres:2019fum}.
The simple pair-counting model is shown by the red line and its uncertainty is shown by the red band. The uncertainty of $\pm 8\%$ is determined by the  difference between the model and  the AV18 calculation for the same $np/pp$  ratio.
The resulting $np/pp$ SRC-pair ratio  uncertainty is shown by the vertical lines.
}
\label{fig:pairCounting}
\end{figure}

As SRC pairs include not only deuteron-like proton-neutron pairs, but also proton-proton and neutron-neutron pairs, the SRC approximation was extended to include all pair types, in what we call the simple pair-counting model~\cite{Nguyen:2020mgo,Li:2022fhh}:
\begin{equation}
    \label{eq:paircounting}
    \begin{split}
    \sigma_{eA}(x_B,Q^2) = K &\cdot  [ N_{np}^A \cdot (\sigma_{ep}+\sigma_{en}) \\
    &+ N_{pp}^A \cdot 2\sigma_{ep} + N_{nn}^A \cdot 2\sigma_{en} ],
    \end{split}
\end{equation}
where $K$ is a kinematic factor and $N_{NN}^A$ is the number of $NN$-SRC pairs (where $NN$ refers to $np$, $pp$, or $nn$)  in a nucleus with mass number $A$.

For the Tritium to Helium-3 cross-section ratio, the simple pair-counting approximation gives:
\begin{equation}
    \label{eq:A3counting}
    \begin{split}
    \frac{\sigma_{T}}{\sigma_{He}} = \frac{N_{np}^{T} \cdot (\sigma_{ep}+\sigma_{en})+N_{nn}^{T} \cdot 2\sigma_{en}}{N_{np}^{He} \cdot (\sigma_{ep}+\sigma_{en})+N_{pp}^{He} \cdot 2\sigma_{ep}}.
    \end{split}
\end{equation}
Using isospin symmetry (i.e., assuming  $N_{nn}^{T} \approx N_{pp}^{He}$ and $N_{np}^{T} \approx N_{np}^{He}$) this simplifies to:
\begin{equation}
    \label{eq:A3counting_iso}
    \begin{split}
    \frac{\sigma_{T}}{\sigma_{He}} = \frac{1 + \frac{\sigma_{ep}}{\sigma_{en}} + 2\frac{N_{pp}}{N_{np}}}{1 + \frac{\sigma_{ep}}{\sigma_{en}} (1+2\frac{N_{pp}}{N_{np}}) },
    \end{split}
\end{equation}
where  we omit the nucleus notation from the number of SRC pairs and always refer to  Helium-3.

While  simplistic, this model relates the SRC-pair isospin structure, $\frac{N_{np}}{N_{pp}}$, to the measured cross-section ratio using known electron-nucleon cross-sections (see solid red line in Fig.~\ref{fig:pairCounting}).
However, the cross section ratio is not very sensitive to the $np/pp$ SRC-pair ratio. A factor of five $np/pp$-ratio variation (from 2 to 10) corresponds to only about $20\%$ change in the cross-section ratio (from $~0.75$ to $~0.95$). Thus, to extract the $np/pp$ SRC-pair ratio from inclusive data, one needs both a small data uncertainty and an equally small model uncertainty.

Inelastic scattering (pion production) and 
rescattering of the outgoing nucleon (Final State Interactions or  FSI) can also impact the cross section and hence the $\frac{N_{np}}{N_{pp}}$ extraction.  Inelastic scattering at $x\ge 1.3$ is negligible~\cite{Benhar:1991af,Benhar:2006wy,Mezzetti:2010by}.
FSI can change the cross-section by up to a factor of $\sim 2$ at high-$x_B$~\cite{frankfurt08b,Mezzetti:2010by}.  However, at high-$x_B$ FSI primarily consists of one nucleon from the SRC pair scattering off the other nucleon in the pair.
Therefore, if the reaction is dominated by interactions with $np$-SRC pairs, then the effects of FSI effect will be the same and will largely cancel in the cross-section ratio for different nuclei.  This cancelation should be even stronger for the cross-section ratio of similar nuclei, e.g., $^3$He to $^3$H.
However, differences in the $np$ and $pp$ (or $nn$) scattering cross-section could lead to different FSI for interactions with $np$- and $pp$-SRC pairs, changing their relative contribution to the measured cross-section and altering the extracted $\frac{N_{np}}{N_{pp}}$ ratio.  This would increase the uncertainty and decrease our confidence in the results of the simple pair-counting model of Ref.~\cite{Li:2022fhh}.

Previous work used Eq.~\ref{eq:A3counting_iso} to extract an $np/pp$ SRC pair ratio of $4.34^{+0.49}_{-0.40}$, quoting a $0.4\%$ model uncertainty attributed to possible isospin symmetry breaking and SRC center-of-mass (c.m.) momentum effects~\cite{Li:2022fhh}.
However, additional model uncertainties can come from the difference between the $pp$ and $np$ pair relative wave functions (due to the tensor-force contribution to the $T=0$, $S=1$ $np$ state~\cite{Weiss:2020mns}), their different FSI contributions, and contributions to the cross-section from interactions with non-SRC states.

To assess these effects
we perform a cross-section ratio calculation using input from \textit{ab initio} nuclear structure calculations.

At high-$Q^2$ the electron scattering cross-section can be approximately factorized as~\cite{kelly96}:
\begin{equation}
    \label{eq:pwia}
    \sigma_{eA} = K' \cdot \sigma_{eN} \cdot S_N(k,E_s),
\end{equation}
where $K'$ is a kinematic factor, $\sigma_{eN}$ is the one-body electron-nucleon off-shell cross-section~\cite{DeForest:1983ahx}, and $S_N(k,E_s)$ is the  spectral function, defined as the probability for finding a nucleon ($N=p,n$) in the nucleus with momentum $k$ and separation energy $E_s$. 
$E_s$ is defined by $E_s \equiv -m_A + m_N + m_{A-1}^*$,
where $m_A$ is the mass of the target nucleus and $m_{A-1}^*$ is the invariant mass of the entire $(A-1)$ system, which may or may not remain intact. 

In the absence of inelastic scattering and final state interactions, the inclusive cross section equals this cross section integrated over all spectral function states ($k$ and $E_s$) within the experimental acceptance and summed over proton and neutron knockout.

This factorized cross-section approximation encapsulates all of the many-body nuclear structure information in the  spectral function. Here we use a spectral function extracted from exact calculations of the three-body ground-state~\cite{AttiKaptari:2005} using the AV18 interaction~\cite{wiringa95} and without irreducible three-body forces.
The spectral function accounts for FSI between the two spectator nucleons but not with the leading nucleon.

\begin{figure}[t]
\centering 
\includegraphics[width=0.9\linewidth]{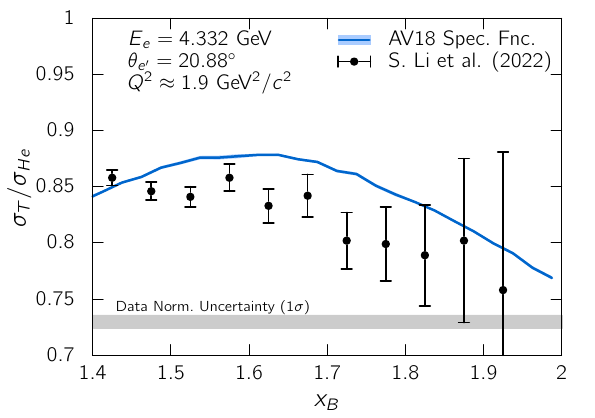}
\caption{
Tritium to Helium-3 $(e,e')$ cross-section ratio, $\sigma_T/\sigma_{He}$, at $Q^2\approx 1.9$ GeV$^2$ as a function of $x_B$ for data~\cite{Li:2022fhh} (black points) and calculation (blue line). 
The factorized cross-section calculation (Eq.~\ref{eq:pwia}) uses an exact \textit{ab initio} spectral function calculated using the AV18 interaction~\cite{AttiKaptari:2005}. The gray band shows the  $1.18\%$ data normalization uncertainty ($1\sigma$).
}
\label{fig:t_over_h}
\end{figure}

Fig.~\ref{fig:t_over_h} shows the results  for the $Q^2=1.9$ GeV$^2$ kinematics of Ref.~\cite{Li:2022fhh}. 
The calculation is in good agreement with the data, showcasing a remarkable success for \textit{ab initio} nuclear structure theory, even at very high-$x_B$.

Fig.~\ref{fig:pairCounting} shows the calculated cross-section ratio  averaged over $1.4 \le x_B \le 1.7$  plotted at  the corresponding $np/pp$ SRC ratio of 11.5, which  was taken from the AV18-based Quantum Monte-Carlo (QMC) calculations of Ref.~\cite{Cruz-Torres:2019fum} (which also included $3N$ forces). For the same $np/pp$ SRC ratio, this calculation differs from the simple pair-counting approximation by $8\%$.

This difference provides a measure of the theoretical uncertainty of the  pair-counting approximation. The impact of this $8\%$ uncertainty is shown in Fig.~\ref{fig:pairCounting} by the red band surrounding the  pair-counting calculation. Despite the  precision of the measured  cross-section ratio, this $8\%$ theoretical interpretation uncertainty leads to a large uncertainty in the extracted $np/pp$ SRC ratio, spanning a factor of $5$ from $2.5$ to $12$.

We  further used the AV18 spectral-function-based cross-section calculation to quantitatively test the claimed SRC-dominance of the  measurements in the plateau region ($1.4\le x_B \le 1.7$) by examining the contributions of low-momentum and low-energy states to the $(e,e')$ cross-section. 

\begin{figure}[t]
\centering 
\includegraphics[width=0.96\linewidth]{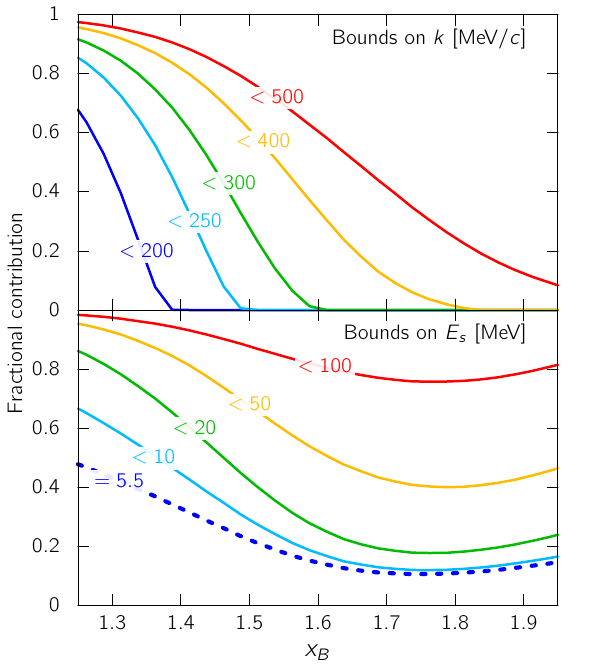}
\caption{
Calculated contributions to the  $^3$He$(e,e')$ cross-section from different initial nucleon states characterized by their initial momentum, $k$, (top) and separation energy, $E_s$ (bottom)~\cite{AttiKaptari:2005}. The dashed blue line in the bottom panel, labeled ``$=5.5$'', represents the two-body $pd$ break-up contribution. Calculations are performed using Eq.~\ref{eq:pwia} for the $Q^2\approx 1.9$ GeV$^2$ kinematics of Ref.~\cite{Li:2022fhh}, using an \textit{ab initio} AV18 spectral function~\cite{AttiKaptari:2005}. 
}
\label{fig:PmEmFractions}
\end{figure}

Figure~\ref{fig:PmEmFractions} shows the fractional contribution to the Helium-3 cross-section as a function of $x_B$ from states with different momentum and energy cutoffs.
While low-momentum contributions are suppressed at  high-$x_B$, there is still significant strength from $k \leq 250$ MeV/c up to $x_B =  1.5$,   i.e., the first two  data points in the SRC-scaling region of  Ref.~\cite{Li:2022fhh}.
More importantly, even at $1.4\le x_B\le 1.7$ there are still significant ($\ge 20\%$) contributions to the cross-section from  high-momentum low-energy nuclear states. These states differ from those typically measured in nucleon-knockout SRC studies, where much higher energies are probed and a clear momentum-energy correlation is observed~\cite{schmidt20}.
Even at  the highest $x_B$, the $^3$He cross-section includes a $\sim10\%$ contribution from two-body breakup (e.g., $(e,e')pd)$ that is clearly not associated with SRC-pair breakup.  These non-SRC-pair states are one source of the discrepancy between the simple pair-counting model and the \textit{ab initio} spectral-function-based calculation.

While the measured inclusive cross-section ratio has some sensitivity to  SRC isospin structure,  the simple pair-counting model ignores the complexities of nuclear structure and therefore has far greater uncertainties than the high-precision data. Thus, we do not see conclusive evidence for the claimed reduction in the  neutron-proton pair dominance in the $A=3$ system.

This conclusion does not preclude a different $np/pp$ SRC-pair ratio in $A=3$ nuclei, resulting from  unique constraints in the $A=3$ system.
The SRC-pair and third-nucleon spins   must sum to the total nuclear spin of 1/2. Similarly, the third nucleon momentum must be equal and opposite to the SRC pair c.m. momentum, potentially limiting the pair formation phase-space and mixing three-nucleon correlation effects~\cite{bagh10}.
These and other  effects could change the $np/pp$ SRC ratio in light-nuclei without requiring a change to our current understanding of high-momentum short-distance interactions.

Last, we  examine the sensitivity of the cross-section ratio to the short-distance $NN$ interaction.
Since exact three-body spectral functions using different $NN$ interaction models are not yet available, we turn to the Generalized Contact Formalism (GCF)~\cite{Weiss:2015mba,Weiss:2016obx,Weiss:2018tbu,Cruz-Torres:2019fum}.

The GCF provides a  SRC-pair based spectral function model that was shown to well reproduce both \textit{ab initio} calculated one- and two-nucleon densities~\cite{Weiss:2016obx,Cruz-Torres:2019fum}, as well as exclusive SRC pair breakup measurements~\cite{Weiss:2018tbu,Duer:2018sxh,schmidt20,Korover:2020lqf,Pybus:2020itv,Patsyuk:2021jea}.
The GCF  approximates the high-momentum high-energy part of the nuclear spectral function as a sum over SRC-pair spectral functions~\cite{Weiss:2018tbu,Pybus:2020itv}:
\begin{equation}
    \label{eq:gcfS}
    \begin{split}
    S_p(k,E_s) = & C_{pn}^{1}S_{pn}^{1}(k,E_s) + C_{pn}^{0}S_{pn}^{0}(k,E_s) + \\
    & 2C_{pp}^{0}S_{pp}^{0}(k,E_s) ,\\
    S_n(k,E_s) = & C_{pn}^{1}S_{pn}^{1}(k,E_s) + C_{pn}^{0}S_{pn}^{0}(k,E_s) + \\
    & 2C_{nn}^{0}S_{nn}^{0}(k,E_s),\\
    \end{split}
\end{equation}
where $S_p$ and $S_n$ are the proton and neutron spectral functions.
$C_{NN}^{\alpha}$ are the nuclear ``contact terms" that specify the number of $NN$ SRC pairs with spin $\alpha$ ($=0, 1$) in nucleus $A$, 
and $S_{NN}^{\alpha}(k,E_s)$ are pair spectral functions given by:
\begin{equation}
    \label{eq:gcfSNN}
    \begin{split}
    S_{NN}^{\alpha}(k_1,E_s) = \frac{1}{4\pi} \int \frac{d^3\mathbf{k_2}}{(2\pi)^3} \; & \left| \phi_{NN}^{\alpha}((\mathbf{k_{rel}}) \right|^2 \\
    &  n_{NN}^{A\alpha}(\mathbf{k_{cm}}) \:\:\: \delta(f(\mathbf{k_2})).
    \end{split}
\end{equation}
$\mathbf{k_{rel}}$ and $\mathbf{k_{cm}}$ are the pair relative and center of mass (c.m.) momenta and $\mathbf{k_1}$ and $\mathbf{k_2}$ are the momenta of nucleons im the SRC pair.
$\phi_{NN}^{\alpha}(\mathbf{k_{rel}})$ are the universal SRC pair relative wave functions, obtained from the zero-energy solution of the Schr{\"o}dinger equation for a given $NN$ interaction model and $n_{NN}^{A\alpha}(\mathbf{k_{cm}})$ is the SRC pair c.m.\ momentum distribution for nucleus $A$, given by a three-dimensional Gaussian~\cite{Cohen:2018gzh} (same for all $NN$ interactions). The delta function $\delta(f(\mathbf{k_2}))$ ensures 4-momentum conservation. See~\cite{Weiss:2018tbu} for details.

While the GCF and the simple pair-counting model both only account for interactions with SRC pairs, the GCF is more realistic by directly accounting for c.m. motion effects and allowing for different $pn$ and $pp$ relative wave functions $\phi_{NN}^{\alpha}$ that are calculated using a specific $NN$ interaction. The corresponding SRC contact terms are extracted from \textit{ab initio} nuclear structure calculations using the same $NN$ interaction model (typically also including $3N$ forces).

\begin{figure}[t]
\centering 
\includegraphics[width=\linewidth]{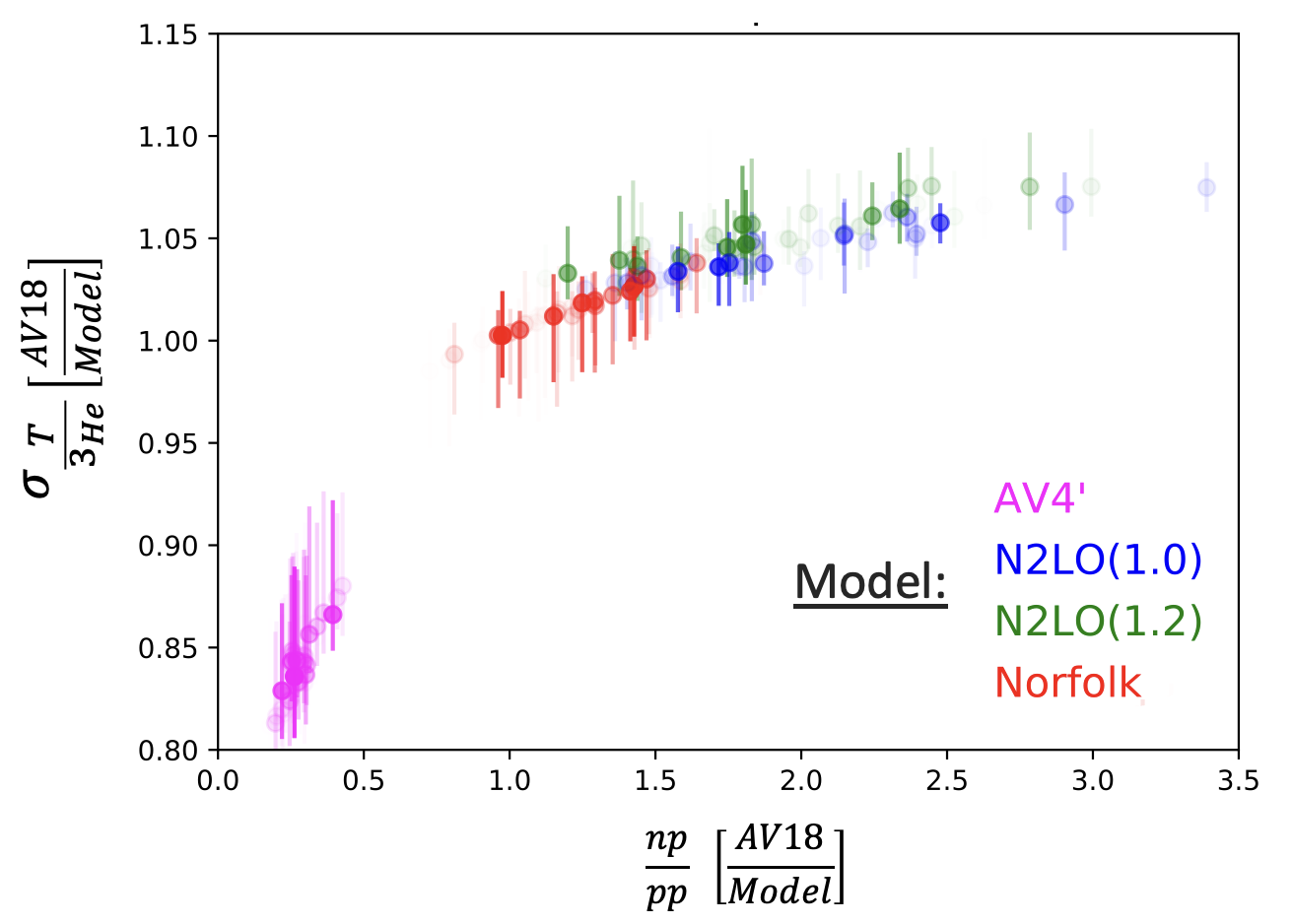}
\caption{Tritium to $^3$He GCF cross-section ratios for different $NN$ interactions (AV4': magenta; Norfolk: red; N2LO(1.0fm): blue; N2LO(1.2fm): green) relative to AV18, plotted versus the corresponding $np/pp$ Contact ratio relative to the AV18 ratio. The different  points represent different input parameters and intensity of the points is proportional to the probability of the input parameters. }
\label{fig:GcfRatios}
\end{figure}

Figure~\ref{fig:GcfRatios} shows the calculated GCF cross-section ratio for a variety of potentials relative to the AV18 calculation. The calculated cross-section ratios are plotted versus the corresponding $np/pp$ Contacts ratio, extracted for the different interactions relative to the AV18 extraction. The calculation follows Ref.~\cite{Weiss:2020mns} using the contact extractions of Ref.~\cite{Cruz-Torres:2019fum}, where different  points represent different samples of the pair c.m.\ momenta and Contact ratios. 
The intensity of the points is proportional to the probability of the input parameters.  The Chiral EFT potentials~\cite{Piarulli:2016vel,Piarulli:2017dwd,Baroni:2018fdn,Gezerlis:2014,Lynn:2016,Lonardoni:2018prc} cross-section ratios agree with the AV18 result to about 5\%. At the same time their corresponding $np/pp$ ratios  vary by a factor of 2.5. The tensor-less AV4' potential~\cite{Wiringa:2002} gives a cross-section ratio that is only 10\% - 15\% smaller than AV18, but an $np/pp$ ratio that is four times smaller. This is another indication that the $A=3$ inclusive cross-section ratio  has little sensitivity to the  SRC-pair isospin structure.

Our previous $A=3$ GCF cross-section ratio calculation using the AV18 interaction (Fig. S1 of Ref.~\cite{Weiss:2020mns}) is about 15\% larger than the exact AV18 spectral-function calculation of Fig.~\ref{fig:t_over_h}.  This difference is likely due to  non-SRC-pair contributions to the spectral function (see Fig.~\ref{fig:PmEmFractions}).
Such contributions are not included in SRC-based approximations, such as the GCF or the simple pair counting model, limiting their applicability for the analysis of inclusive cross-sections.

%-----------
To conclude,  Ref.~\cite{Li:2022fhh} measured the $^3$He/$^3$H inclusive $(e,e')$ cross-section ratio, compared it to a simple pair-counting model, and interpreted the difference as evidence for ``unexpected structure in the high-momentum wavefunction for hydrogen-3 and helium-3''

In contrast, we showed that these ratios agree with a theoretical calculation using a full \textit{ab initio} AV18-interaction-based three-body spectral function (Fig.~\ref{fig:t_over_h}). This suggests that the data are consistent with current nuclear theory and contradicts the claim of Ref.~\cite{Li:2022fhh}. 

The \textit{ab initio}  calculation differs from the simple pair-counting model by $8\%$ (Fig.~\ref{fig:pairCounting}) and from the less-simple GCF pair-counting model for the same $NN$ interaction by 15\%. These differences appear to stem from one- and three-nucleon contributions to the claimed  SRC-pair scaling region (Fig.~\ref{fig:PmEmFractions}).  This difference is  amplified by the small sensitivity of the measured inclusive cross-section ratio to the underlying SRC pair isospin structure, leading to a factor of five uncertainty in the $np/pp$ ratio. 

 Using the GCF we found that the cross-section ratio is not sensitive to the $NN$-interaction model  (see Fig.~\ref{fig:GcfRatios}) and hence not sensitive to the $np/pp$ SRC-pair ratio. This is in contrast with  one- and two-nucleon knockout cross-section ratios that are very sensitive to the $NN$-interaction model.

Therefore, current inclusive cross-section ratios are consistent with our knowledge of the nuclear structure of the $A=3$ system  and cannot be used to precisely quantify  the $np/pp$ SRC-pair ratio. The overarching quest to fully understand the short-distance structure of the $A=3$ system requires additional work through varied experimental measurements and state-of-the-art theoretical calculations with quantified uncertainties.

\begin{acknowledgments}
We thank C. Ciofi degli Atti and L. Kaptari for the $A=3$ spectral function calculations. We also thank N. Rocco and A. Lovato for valuable discussions. This work was supported by the U.S. Department of Energy (DOE) Grant Nos. DE-SC0016583, DE-SC0020240, DE-FG02-96ER40960, DE–FG02–93ER40771, and DE-FG02-97ER-41014, the Israeli Science Foundation grant No. 1086/21, and Pazy Foundation. R.W.\ was supported by the Laboratory Directed Research and Development program of Los Alamos National Laboratory under project number 20210763PRD1.
\end{acknowledgments}

%%%%%%%%%%%%%%%%%%%%%%%%%%%%%%%%%%%%%%%%%%%%%%%%%%%%

%\bibliographystyle{unsrt}
\bibliography{references}

%%%%%%%%%%%%%%%%%%%%%%%%%%%%%%%%%%%%%%%%%%%%%%%%%%%%

\end{document}